# PRESENT STATUS OF MEDICAL PHYSICS PRACTICE IN MEXICO: AN OCCUPATIONAL ANALYSIS

D. Garcia-Hernandez[1], X. Lopez-Rendon[2], M. Hernandez-Bojorquez[3], J. A. Herrera-Gonzalez[4], O. E. Soberanis-Dominguez[5], S. A. Gonzalez-Azcorra[6], J. P. Cruz-Bastida[7]

[1]UNEME de Oncologia, Servicios de Salud de Zacatecas, Mexico
[2]Departmento de Neuroimagen, Instituto Nacional de Neurologia y Neurocirugia, Mexico City, Mexico
[3]Departmento de Radioterapia, The American British Cowdray Medical Center, Mexico City, Mexico
[4]Unidad de Radioneurocirugia, Instituto Nacional de Neurologia y Neurocirugia, Mexico City, Mexico
[5]Departmento de Medicina Nuclear, Centro Oncologico Privado, Merida, Mexico
[6]Departamento de Radioncologia, Hospital Agustin O'Horan, Merida, Mexico
[7]Department of Radiology, University of Chicago, Chicago, USA

*Abstract*— The clinical practice of Medical Physics in Mexico has not been subject of comprehensive occupational analyses. The absence of such studies not only arises radiation safety concerns, but also imposes challenges to work-policy making. This work presents an initial effort to overview the current occupational status of clinical Medical Physics in Mexico. Our motivation and final goal is to support, based on data, the legal recognition of Medical Physics high-end training, and to provide information that will potentially improve the Mexican health-care system. For the ease of analysis, the concept of "person(s) developing Medical Physics tasks" (PDMPT) is introduced to refer to professionals playing clinical medical physicist's (cMP) roles, disregarding academic profile or training. A database of PDMPT in Mexico was built from official sources and personal communication with peers. Our database included: employer(s), specialty and academic profile. It was found that 133 hospitals in Mexico employ PDMPT, 49% of which are public institutions. A total of 360 positions involving cMP roles were identified in the National Health-Care System, 77% of which corresponded to radiation oncology. Public health services hold 65% of the reported positions. Cases of double- and triple-shift workers where identified in this study, as 283 PDMPT occupied the 360 reported positions. Of all PDMPT, 32% were women. Only 40% of PDMPT hold a graduate degree in Medical Physics, 46% of which were located in the most densely populated region of Mexico. Our data suggests that Mexico is far from fulfilling the international recommendations regarding cMP academic profile; however, this problem could be solved in the near future for the specific cases of radiation oncology and nuclear medicine services in the public health-care sector.

*Keywords*— Mexico, medical physicists, manpower and services, professional qualifications, working conditions, demographics, medical physicist's education.

I. INTRODUCTION

To the best of our knowledge, the practice of Medical Physics in Mexico has not been subject of a comprehensive, systematic and rigorous occupational analysis. Even though the International Atomic Energy Agency (IAEA) has made efforts to make public directories of radiation therapy [1] and nuclear medicine [2] centers worldwide, such resources are based on voluntary contribution and do not include occupational statistics. Previous works [3,4] have also provided some insight on the status of Medical Physics practice in Mexico; however, such studies have been mainly focused on the analysis of radiation therapy centers and can be considered outdated.

The absence of an updated occupational database of the Medical Physics practice in Mexico not only arises radiation safety concerns, but also imposes challenges to work-policy making. Particularly, it has become challenging to state the definition, roles and responsibilities of clinical medical physicist (cMP) in legal terms and practice. Ideally, the definition of cMP would only include individuals with the academic profile and clinical training suggested by international organizations [5-8]; however, in Mexico, on-job empirical training is often considered sufficient to bypass international recommendations, arising ambiguities on the definition of cMP. For the sake of discussion, in this work, professionals playing cMP roles are referred to as "person(s) developing medical-physics tasks" (PDMPT), disregarding academic profile or training.

Mexican legal frameworks acknowledge the figure of "Medical Physicist" [*sic*] since 2011 [9,10,11]; nonetheless, the definition of such figure is significantly different from international recommendations and subject to personal interpretations. While graduate Medical Physics studies are strongly encouraged by most international organizations, Mexican regulations accept as a valid academic profile a bachelor's degree in Physics or "related engineering degrees" [*sic*]. In addition, the responsibilities of a cMP in the Mexican health-care system are unjustifiably bounded. Up to now, the presence of a cMP is only mandatory for radiation therapy facilities and nuclear medicine services in use of "complex" [*sic*] imaging equipment [9].

A reason why PDMPT profile often fails to meet international recommendations is a limited academic development of the country. In Mexico, graduate Medical Physics programs are relatively recent and small in number and size; to date, only two graduate Medical Physics programs are offered nationwide [12,13], from which 225 students have completed a master's degree. Moreover, just about half of the Medical Physics graduates are employed as PDMPT, while the others opt for academia, industry, etc. Recently, Medical Physics undergraduate programs and specializations have emerged in the country, but these academic alternatives have not proven to be equivalent to career paths suggested by international organizations; in addition, the emergence of undergraduate Medical Physics programs further introduces ambiguities in the definition of cMP and could incite unfair competition among PDMPT.

For the reasons exposed above, it is of major importance to determine the current occupational status of Medical Physics practice in Mexico. This work presents an initial effort to conduct such task. Our motivations and final goals are to support, based on data, the legal recognition of Medical Physics high-end training, and to provide information that will potentially improve the Mexican health-care system.

## II. Background

For the ease of reading, and to help international audiences better understand the contents in this work, Mexican demographics and other relevant descriptions are provided in this section.

As of 2017, Mexico is estimated to have a population of 123.6 million [14] distributed in 32 states. The most densely populated states are Mexico City (CDMX) and the State of Mexico (EdoMex). Together, these states concentrate approximately 26 million people; that is, about 21% of the total population.

An estimate of 35,398 centers constitute the Mexican health-care network [15], including: hospitals, ambulatory-care clinics and ancillary-service providers (for example, diagnostic and clinical laboratories). These centers are classified as either public or private. Public health-care facilities in Mexico belong, mainly, to the following providers: SSA, IMSS, ISSSTE, PEMEX, SEDENA and SEMAR (see Table 1).

By law, the SSA coordinates the whole National Health-Care System, for both private and public institutions.

*Table 1: Mexican institutions of relevance for this paper.*

| Acronym | Institution | Description |
|---------|-------------|-------------|
| SSA | Secretaria de Salud | Secretary of public health |
| ISSSTE | Instituto de Seguridad y Servicios Sociales de los Trabajadores del Estado | Social security system for federal workers |
| IMSS | Instituto Mexicano del Seguro Social | Social security system for workers in the private sector |
| PEMEX | Petroleos Mexicanos | Social security system for workers of the Mexican state-owned petroleum company |
| SEDENA | Secretaria de la Defensa Nacional | Social security for Army and Air Force members |
| SEMAR | Secretaria de Marina | Social security for Navy members |
| CNSNS | Comision Nacional de Seguridad Nuclear y Salvaguardias | Nuclear Safety authority |
| COFEPRIS | Comision Federal para la Proteccion contra Riesgos Sanitarios | Regulatory body for protection against sanitary risks |
| SHCP | Secretaria de Hacienda y Credito Publico | Secretary of Treasury |

## III. Methods

A database of PDMPT in Mexico was created by combining three major sources: (i) alumni follow-up databases of the existing Medical Physics graduate programs [12,13]; (ii) personal communication with colleagues around the country; (iii) official public-access information provided by the Mexican government (by request) [16]. The information provided by the Mexican government included reports from five different dependencies: SSA, CNSNS, IMSS, ISSSTE and SHCP (see Table 1). The resulting database was last updated in April 2019.

The database created in this work included the following information about PDMPT: (i) place(s) of employment, (ii) Medical Physics specialty, and (iii) academic profile. Medical Physics specialties were classified as follows: Radiation Oncology (RT), Nuclear Medicine (NM) and Diagnostic and Interventional Radiology (D&IR). To provide context to our results, the number of physicians, imaging/treatment devices and health-care centers accompany our statistics. The number of physicians per specialty was obtained from Ref. [17], while the number of imaging/treatment devices (hereafter referred to as "equipment") was obtained from SSA database. For D&IR, the number of equipment included: CT scanners, mammography units, angiography units and MR scanners. For NM, only PET-CT, PET-MRI and SPECT-CT scanners were taken into account. Besides hospitals, the reported number of imaging centers includes mobile units, small non-specialized clinics, and for-profit imaging laboratories. Monthly income statistics of PDMPT were also gathered as part of our study, when available.

## IV. Results

### A. Health-care Centers Employing PDMPT

A total of 133 centers employing PDMPT were identified across the nation (mainly hospitals); 35 of them were located in the most densely populated region of Mexico (CDMX + EdoMex). Figure 1 shows the number of centers employing at least one PDMPT for each state of the country. The number of public centers on each state is also depicted in the figure. Nationwide, almost half (49%) of the total number of centers are public.

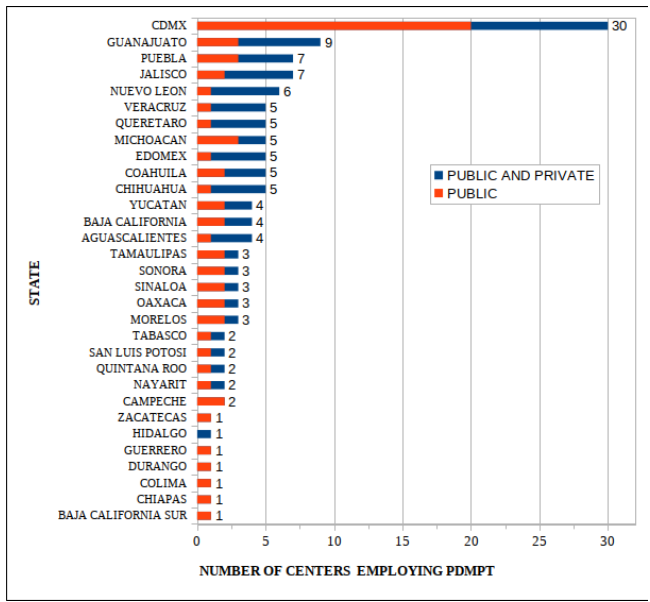

*Figure 1: The number of centers employing PDMPT is shown for each state of the Mexican Republic. The number of public centers per state is also depicted.*

### B. Health-care Centers' Licenses and Equipment

According to information provided by the CNSNS, 106 institutions are licensed to practice teletherapy in Mexico, 52 of them public; this number is consistent with the 164 treatment devices ($^{60}$Co units and linacs) registered by SSA. As for brachytherapy, 72 licensed centers (47 of them public) concentrate 81 treatment devices (x-ray generators, remote-afterloading systems with $^{192}$Ir and $^{60}$Co sources, and manual-loading $^{137}$Cs, $^{90}$Y and $^{125}$I sources). In total 109 institutions are licensed to practice either tele- or brachy- therapy. The CNSNS also reports 177 institutions with nuclear medicine licenses, 49 of them public. According to SSA, 55 centers (23 of them public) are in use of hybrid imaging equipment (71 devices in total). In the case of D&IR, SSA reports 793 CT scanners, 1473 mammography units, 187 angiography units and 316 MR scanners.

### C. PDMPT Occupational Status

A total of 360 positions involving cMP roles (hereafter referred to as "cMP positions", for simplicity) were identified in both public and private institutions. Such positions are occupied by 283 PDMPT, and their distribution in the National Health-Care System is depicted in Figure 2. Note that this figure does not include PEMEX hospitals, as there are not any records of PDMPT hired by such institution. It is worth mentioning that 32% of all PDMPT are women, and that the most densely populated region of Mexico (CDMX + EdoMex) concentrates 42% of PDMPT.

The public health-care system holds 235 of all the cMP positions (65%), which are occupied by 216 PDMPT. The geographical distribution of such positions is very similar to that of the whole health-care system (Figure 1).

Most PDMPT are employed under a variety of inconsistent job codes. Even though IMSS and SSA have recently introduced a job code labeled as "Medical Physicist", such designation is currently underutilized. For example, in SSA only about 10% of PDMPT are employed as "Medical Physicist". In addition, the "Medical Physicist" job code at IMSS is divided into sub-codes, which further introduces confounding factors when attempting to rigorously define a cMP.

### D. PDMPT Academic Profiles

PDMPT education widely differs, ranging from technical degrees to postdoctoral training. According to CNSNS records, there are 5 cases of PDMPT without a registered bachelor's degree. According to our database, 40% of all PDMPT hold a graduate degree in Medical Physics, 46% of which are employed in CDMX + EdoMex. Table 2 shows in detail the geographical distribution of PDMPT holding graduate degrees; data for public institutions is given in separate columns. It is worth emphasizing that only 6 of all 283 PDMPT (~2%) have achieved partial certification from the International Medical Physics Certification Board [18], five of which are women.

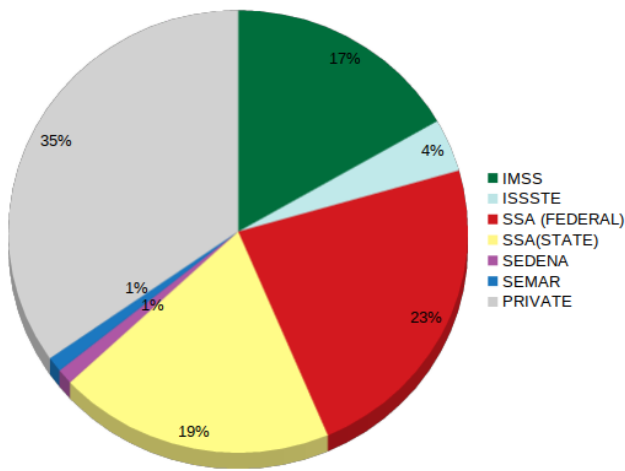

Figure 2: Distribution of cMP positions in the Mexican health-care system. SSA centers were divided into federal- and state- providers.

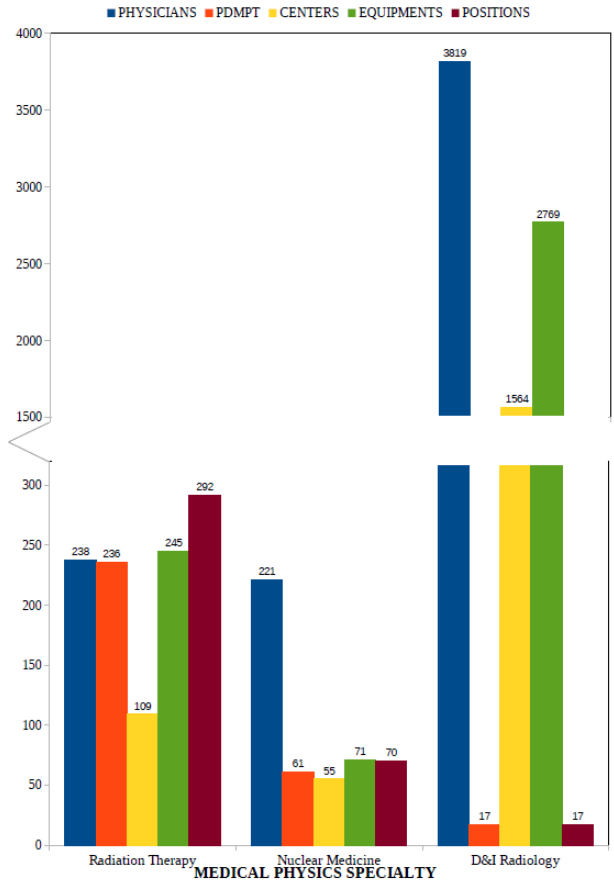

Figure 3: Manpower, facilities and equipment per Medical Physics specialty in Mexico.

### E. PDMPT Specialization

Figure 3 shows the number of PDMPT per specialty nationwide. This figure also depicts the number of centers, equipment, physicians, and cMP positions associated to each specialty, as reference. Note that, of all cMP positions, 77% correspond to RT, 18% correspond to NM and only 5% correspond to D&IR. Large discrepancies were found when comparing this data with previous works [3,4] and IAEA databases [1,2], which suggests that such sources may need to be updated. It is worth emphasizing that, with exception of RT services, PDMPT are considerably outnumbered by physicians.

According to our study, 11% of all PDMPT practice more than one of the specialties listed above (RT, NM or D&IR). Also, in 19 cases, the PDMPT were required by the employer to perform intra-center chores in more than one specialty.

### F. PDMPT Income

Even though income statistics were not always available, rough estimates can be made from our data. In the case of public institutions, PDFLM annual income approximately ranges from $11k to $24k USD. In contrast, the annual income in private institutions approximately ranges from $14k to $42k USD. The most typical annual income (mode) was estimated to be around $14k USD.

### V. Discussion and Conclusions

From the previously described data, lessons, conclusions and future challenges immediately arise. First, it becomes evident that the number of PDMPT in Mexico is insufficient. The most alarming case corresponds to D&IR services, where the number of centers and equipment is several times larger than the associated PDMPT. A less severe difference, but still subject of concern, was found in NM services. Only RT services count with approximately one PDMPT per equipment. This may be due to the fact that, traditionally, Mexican regulations for RT are subject to more severe inspections and verifications of compliance, compared to other specialties.

There is a considerable number of cases of PDMPT with double or triple shifts that need to be revised. Double- and triple-shift workers may be subject of radiation safety concerns, as the burn-out phenomenon [19] could negatively impact their performance. PDMPT whose multiple appointments encompass more than one specialty arise additional concerns; in such cases, PDMPT competence over time is potentially challenged by the rapid and overwhelming development of all specialties in Medical Physics. In order to reduce the incidence of multi-shifting, the high demand of PDMPT in the Mexican health-care system needs to be alleviated, and the labor conditions of currently employed PDMPT need to be examined; precarious job conditions of PDMPT may be a cause of multi-shifting.

Our results show a clear gender imparity. The explanation behind this finding is beyond the scope of this work; nonetheless, the implementation of gender policies in academia and the Mexican health-care system could help to balance this trend.

Another result that rises social concerns is the geographical distribution of PDMPT, as almost half of the available manpower is in service of one fifth of the Mexican population (CDMX + EdoMex). Furthermore, the most highly qualified portion of PDMPT (graduate degree holders) is mainly distributed in CDMX + EdoMex.

Currently, the number of PDMPT with an adequate academic profile [5-8] is insufficient to cover the cMP demand of the Mexican health-care system; however, the following scenario must be considered. Setting aside D&IR services (where the employment of external consulting could be argued in some cases), it is not unrealistic to pursue the employment of professionals with graduate studies in Medical Physics to cover the public sector cMP positions in the near future. As a matter of fact, the total number of alumni of the available medical physics graduate programs is already comparable to the number of licenced RT and NM centers. By doing so, the Mexican health-care system would significantly strengthen and set an important precedent in Latin America. In addition, such a measure would eliminate any potential risk of unfair competition among PDMPT. But first, to make this possible, a rigorous definition of cMP, based on international standards, needs to be recognized by Mexican authorities. Other measures that could help to achieve this goal include: to create Medical Physics graduate programs nationwide, to reinforce the existing Medical Physics graduate programs, and to explain students that some career paths do not provide adequate qualifications to perform as a cMP.

International board certification of PDMPT, on the other hand, is far from being a reality in the short term according to our statistics. Personal communication with graduate degree holders suggested an unwillingness to pursue this status, mainly due to the fact that actual legislation does not contemplate certification as a requirement to practice.

To the best of our knowledge, this is the first work that presents a comprehensive occupational study of PDMPT in Mexico. As discussed above, this kind of studies can help to identify important issues in current Medical Physics practice; therefore, it is desirable to periodically update and further extend our database. While the Mexican government's transparency law enables access to resources of major relevance for this purpose, requesting such materials and filling gaps in their contents is often time consuming for independent researchers. From our experience, three major challenges arise when doing occupational studies solely based on official reports: (i) a continuous and rapid PDMPT inter-center mobility; (ii) reluctance of employers to report PDMPT enrollment, due to legal implications; (iii) lack of adequate job codes to employ PDMPT in some centers. It is therefore reasonable to believe that occupational studies would be more efficient if conducted by special governmental commissions.

### Acknowledgments

We would like to thank friends and colleagues who kindly provided information that helped create and cross-check the database presented in this work. Our special recognition to INAI for safeguarding the human right to access information.

### Disclaimer

The views and declarations expressed in this article are those of the authors and do not necessarily reflect the official policy or position of their associated institutions.

| | CENTERS WITH PDMPT | | | POSITIONS | PDMPT | | | | POSITIONS | PUBLIC SECTOR | | |
| | | | | | | | | | | | PDMPT | | |
| | TOTAL | PUBLIC | PRIVATE | | TOTAL | POSTGRAD. | MED. PHYS. POSTGRAD. | P. CERTIFIED | | TOTAL | POSTGRAD. | MED. PHYS. POSTGRAD. |
|---|---|---|---|---|---|---|---|---|---|---|---|---|
| AGUASCALIENTES | 4 | 1 | 3 | 5 | 4 | 2 | 2 | 2 | 2 | 2 | 1 | 1 |
| BAJA CALIFORNIA | 4 | 2 | 2 | 6 | 5 | 3 | 1 | - | 4 | 4 | 2 | 1 |
| BAJA CALIFORNIA SUR | 1 | 1 | - | 2 | 2 | 2 | 2 | - | 2 | 2 | 2 | 2 |
| CAMPECHE | 2 | 2 | - | 4 | 4 | 1 | - | - | 4 | 4 | 1 | - |
| CHIAPAS | 1 | 1 | - | 1 | 1 | 1 | 1 | - | 1 | 1 | 1 | 1 |
| CHIHUAHUA | 5 | 1 | 4 | 7 | 5 | 1 | 1 | - | 2 | 2 | - | - |
| COAHUILA | 5 | 2 | 3 | 10 | 8 | 3 | 2 | - | 5 | 5 | 3 | 2 |
| COLIMA | 1 | 1 | - | 2 | 2 | 2 | 2 | - | 2 | 2 | 2 | 2 |
| DURANGO | 1 | 1 | - | 2 | 2 | 1 | - | - | 2 | 2 | 1 | - |
| GUANAJUATO* | 9 | 3 | 6 | 18 | 13 | 8 | 6 | - | 10 | 9 | 7 | 5 |
| GUERRERO | 1 | 1 | - | 1 | 1 | 1 | 1 | - | 1 | 1 | 1 | 1 |
| HIDALGO* | 1 | - | 1 | 3 | 3 | 3 | 2 | - | - | - | - | - |
| JALISCO | 7 | 2 | 5 | 23 | 21 | 8 | 6 | - | 16 | 16 | 5 | 3 |
| MICHOACAN | 5 | 3 | 2 | 9 | 6 | 2 | 1 | - | 5 | 4 | 2 | 1 |
| MORELOS* | 3 | 2 | 1 | 8 | 6 | 4 | 4 | - | 6 | 5 | 4 | 4 |
| NAYARIT | 2 | 1 | 1 | 4 | 3 | 3 | 3 | - | 3 | 3 | 3 | 3 |
| NUEVO LEON | 6 | 1 | 5 | 17 | 11 | 5 | 4 | - | 4 | 4 | 1 | 1 |
| OAXACA | 3 | 2 | 1 | 7 | 6 | 4 | 4 | - | 6 | 6 | 4 | 4 |
| PUEBLA | 7 | 3 | 4 | 17 | 14 | 9 | 4 | - | 11 | 10 | 6 | 3 |
| QUERETARO* | 5 | 1 | 4 | 8 | 7 | 6 | 5 | - | 3 | 3 | 2 | 1 |
| QUINTANA ROO | 2 | 1 | 1 | 2 | 2 | - | - | - | 1 | 1 | - | - |
| SAN LUIS POTOSI | 2 | 1 | 1 | 5 | 3 | 2 | 1 | - | 3 | 3 | 2 | 1 |
| SINALOA | 3 | 2 | 1 | 5 | 4 | 4 | 3 | - | 4 | 4 | 4 | 3 |
| SONORA | 3 | 2 | 1 | 9 | 9 | 2 | 1 | - | 7 | 7 | 2 | 1 |
| TABASCO | 2 | 1 | 1 | 5 | 3 | 2 | 1 | - | 3 | 3 | 2 | 1 |
| TAMAULIPAS | 3 | 2 | 1 | 5 | 5 | 1 | 1 | - | 2 | 2 | 1 | 1 |
| TLAXCALA | - | - | - | - | - | - | - | - | - | - | - | - |
| VERACRUZ | 5 | 1 | 4 | 8 | 5 | 3 | 2 | - | 3 | 3 | 2 | 2 |
| YUCATAN | 4 | 2 | 2 | 14 | 13 | 4 | 3 | - | 11 | 11 | 3 | 2 |
| ZACATECAS | 1 | 1 | - | 3 | 3 | 3 | 2 | - | 3 | 3 | 3 | 2 |
| CDMX+EDOMEX* | 35 | 21 | 14 | 150 | 118 | 67 | 52 | 4 | 109 | 95 | 55 | 43 |
| MEXICO | 133 | 65 | 68 | 360 | 283 | 153 | 114 | 6 | 235 | 216 | 122 | 91 |

Table 2. Geographical distribution and academic profile of PDMPT employed in Mexico. Data for public centers is presented in separate columns. The table notation is specified as follows: POSTGRAD.= graduate degree in any field; MED. PHYS. POSTGRAD.= graduate degree in Medical Physics; P. CERTIFIED= partially certified by IMPCB. Rows marked with (*) include cases of PDMPT working in two different states (a total of 6).

Contact of the corresponding author:

Author: Juan P. Cruz-Bastida, Ph. D. (Medical Physics)
Institute: Department of Radiology, University of Chicago
Street: 5841 South Maryland Avenue, MC2026
City: Chicago, IL
Country: USA
Email: cruzbastida@uchicago.edu